\documentclass[journal]{IEEEtran}
\usepackage[utf8]{inputenc}
\usepackage{bm}
\usepackage{epsfig,amsfonts,amsbsy,bm,mathrsfs}
\usepackage{amssymb,amsmath,amsthm,latexsym,amscd,amsfonts}
\usepackage{lettrine}
\usepackage{authblk}
\usepackage{graphics}
\usepackage{graphicx}
\usepackage{psfrag,float}
\usepackage{pstricks}
\usepackage{pst-plot}
\usepackage{cite}
\usepackage{balance}
\usepackage{epsfig}
\usepackage{epstopdf}
\usepackage{bbm}
\usepackage{enumitem}
\usepackage{dsfont}
\usepackage{setspace}
\usepackage{float}
\usepackage{relsize}
\usepackage{comment}
\usepackage{subfigure} 
\usepackage{algorithm}
\usepackage[noend]{algpseudocode}
\usepackage[nolist]{acronym}
\usepackage{tabularx}
\usepackage{pifont}
\usepackage{units}

\usepackage{tikz}
\usetikzlibrary{calc}
\usepackage{xcolor}
\usepackage{tabularx}
\usepackage{colortbl}
\usepackage{pgfplots}
\pgfplotsset{compat=newest}
\usetikzlibrary{plotmarks}
\usetikzlibrary{arrows.meta}
\usepgfplotslibrary{patchplots}
\usepackage{grffile}
\newlength\fheight 
\newlength\fwidth 
\usepgfplotslibrary{fillbetween}

\makeatletter
\newcommand{\gettikzxy}[3]{%
  \tikz@scan@one@point\pgfutil@firstofone#1\relax
  \edef#2{\the\pgf@x}%
  \edef#3{\the\pgf@y}%
}
\definecolor{mygray}{gray}{0.6}

\begin{acronym}[ACRONYM]
\acro{AI}{artificial intelligence}
\acro{AoA}{angle-of-arrival}
\acro{AoD}{angle-of-departure}
\acro{BS}{base station}
\acro{BP}{belief propagation}
\acro{CDF}{cumulative density function}
\acro{CFO}{carrier frequency offset}
\acro{CRB}{Cram\'er-Rao bound}
\acro{DA}{data association}
\acro{D-MIMO}{distributed multiple-input multiple-output}
\acro{DL}{downlink}
\acro{EM}{electromagnetic}
\acro{FIM}{Fisher information matrix}
\acro{GDOP}{geometric dilution of precision}
\acro{GNSS}{global navigation satellite system}
\acro{GPS}{global positioning system}
\acro{IP}{incidence point}
\acro{IQ}{in-phase and quadrature}
\acro{ISAC}{integrated sensing and communication}
\acro{ICI}{inter-carrier interference}
\acro{JCS}{Joint Communication and Sensing}
\acro{JRC}{joint radar and communication}
\acro{JRC2LS}{joint radar communication, computation, localization, and sensing}
\acro{IMU}{inertial measurement unit}
\acro{IOO}{indoor open office}
\acro{IoT}{Internet of Things}
\acro{IRN}{infrastructure reference node}
\acro{KPI}{key performance indicator}
\acro{LoS}{line-of-sight}
\acro{LS}{least-squares}
\acro{MCRB}{misspecified Cram\'er-Rao bound}
\acro{MIMO}{multiple-input multiple-output}
\acro{ML}{maximum likelihood}
\acro{mmWave}{millimeter-wave}
\acro{NLoS}{non-line-of-sight}
\acro{NR}{new radio}
\acro{OFDM}{orthogonal frequency-division multiplexing}
\acro{OTFS}{orthogonal time-frequency-space}
\acro{OEB}{orientation error bound}
\acro{PEB}{position error bound}
\acro{VEB}{velocity error bound}
\acro{PRS}{positioning reference signal}
\acro{QoS}{Quality of Service}
\acro{RAN}{radio access network}
\acro{RAT}{radio access technology}
\acro{RCS}{radar cross section}
\acro{RedCap}{reduced capacity}
\acro{RF}{radio frequency}
\acro{RIS}{reconfigurable intelligent surface}
\acro{RFS}{random finite set}
\acro{RMSE}{root mean squared error}
\acro{RTK}{real-time kinematic}
\acro{RTT}{round-trip-time}
\acro{SLAM}{simultaneous localization and mapping}
\acro{SLAT}{simultaneous localization and tracking}
\acro{SNR}{signal-to-noise ratio}
\acro{ToA}{time-of-arrival}
\acro{TDoA}{time-difference-of-arrival}
\acro{TR}{time-reversal}
\acro{TXRX}[TX/RX]{transmitter/receiver}
\acro{Tx}{transmitter}
\acro{Rx}{receiver}
\acro{UE}{user equipment}
\acro{UL}{uplink}
\acro{UWB}{ultra wideband}
\acro{XL-MIMO}{extra-large MIMO}
\end{acronym}
\begin{document}
\bstctlcite{IEEEexample:BSTcontrol}
\title{Radio Localization and Sensing -- Part I: Fundamentals}
\specialpapernotice{(Invited Paper)}
\author{Henk Wymeersch,~\IEEEmembership{Senior Member,~IEEE}, 
        Gonzalo Seco-Granados,~\IEEEmembership{Senior Member,~IEEE}
\thanks{
This work was supported by the European Commission through the H2020 project Hexa-X (Grant Agreement no.~101015956),by the ICREA Academia Program, and by the Spanish R+D project PID2020-118984GB-I00.}
\thanks{Henk Wymeersch is with the Department
of Electrical Engineering, Chalmers University of Technology, 41258 Gothenburg, Sweden (e-mail: henkw@chalmers.se). 
Gonzalo Seco-Granados is with the Department of Telecommunications and Systems Engineering, Universitat Autonoma de Barcelona, 08193 Bellaterra, Barcelona, Spain (e-mail: gonzalo.seco@uab.cat).}}

\setlength{\abovedisplayskip}{4pt}
\setlength{\belowdisplayskip}{4pt}
\maketitle
\begin{abstract}
This letter is part of a two-letter tutorial on radio localization and sensing, with focus on mobile radio systems in 5G mmWave and beyond. Part I introduces the fundamentals, covering an overview of the relevant literature, as well as the different aspects of  localization and sensing problems. Then, different performance metrics are presented, which are important in the evaluation of methods. Methods are detailed in the last part of this letter. Part I thus provides the necessary background to delve into more forward-looking problems in Part II. 

\end{abstract}
\begin{IEEEkeywords}
Localization, sensing, orientation estimation, synchronization.
\end{IEEEkeywords}
\vspace{-5mm}
\section{Introduction}
\begin{figure}
    \centering
    \includegraphics[width=1\columnwidth]{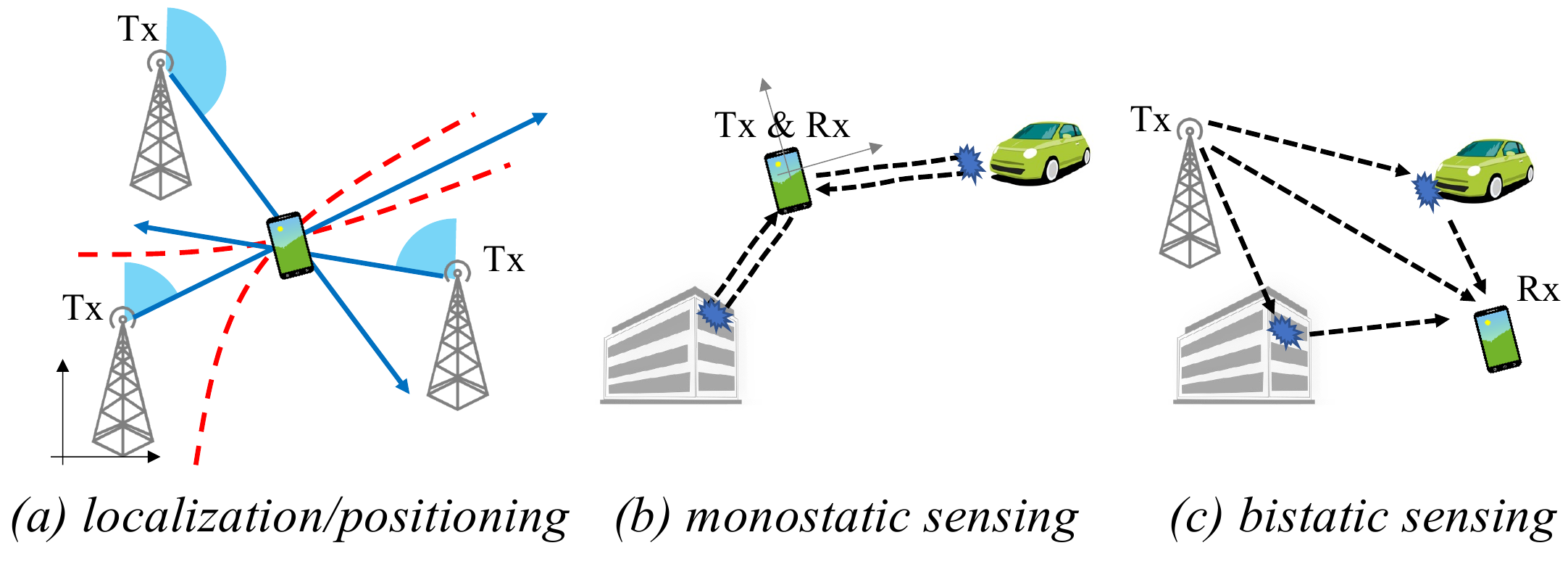}
    \vspace{-8mm}
    \caption{ {(a) Downlink localization of a connected \ac{UE}; (b) Monostatic sensing where a \ac{UE} acts as a {radar} sensor, with co-located transmitter and receiver; (c) bistatic sensing example with \ac{BS} transmitter and \ac{UE} receiver. }}
    \label{fig:locvssense}
    \vspace{-6mm}
\end{figure}

\PARstart{L}{ocalization} {(a term  from  robotics~\cite{ThrunProbRobots05}) and positioning (a term from  navigation and radio communication~\cite{gustafsson2005mobile}) will be interchangeably used for estimation of the state (position, orientation) of a connected device in a global frame of reference (see Fig.~\ref{fig:locvssense}-(a)). 
\emph{Sensing} is broader and 
covers everything from channel parameter estimation and carrier sensing to presence detection~\cite{chaccour2022seven}. In this letter, sensing will refer to state estimation of a passive object in the frame of reference of the sensor (see Fig.~\ref{fig:locvssense}-(b,c)), and thus includes  radar~\cite{patole2017automotive} and device-free localization~\cite{shastri2022review}.}


Radio localization stems from  military satellite-based navigation systems, most notably the \ac{GPS}~\cite{kaplan2017understanding}. 
When a receiver estimates pseudo-ranges from at least 4 synchronized satellites with a favorable geometric configuration, it can determine its 3D position and clock bias. Performance is mainly limited by signal blockages and multipath reflections. Positioning has also been part of evolving cellular standards~\cite{SurveyCellularRadioLocalization--Rosado_others_G.Seco-Granados}. Modern communication systems rely on a combination of time and angle measurements from several \acp{BS} based on dedicated pilot resources to determine the 3D position of a \ac{UE}~\cite{DwivediCOMMAG2021}. 

Radar sensing also has roots in the military, with  surveillance radar systems during World War II providing early warning of incoming bombers~\cite[Vol. III]{richards2010principles}. Due to its myriad of applications, radar has seen enormous developments, e.g., 
in automotive applications, 
where a modern radar can detect and track tens of  moving objects and determine their distance/range, angle, and radial velocity, in the frame of reference of the radar, with very high accuracy~\cite{bilik2019rise}. In contrast to localization, monostatic radar is a local process, and can thus rely on a tailored, highly specialized, and hardware friendly waveforms, without strict standardization constraints. Bistatic sensing, on the other hand, is similar to communication and localization (i.e., when the transmitter or receiver have an unknown position). 
With radar systems and communication systems expected to operate in similar frequency bands, 
there is a potential convergence, 
both in terms of hardware and signals, of sensing and communication systems~\cite{ma2020joint,barneto2021full}. Such \ac{ISAC}, in addition to highly accurate 6D positioning, is expected to be among the main features of 6G~\cite{chaccour2022seven}. 

This letter introduces the fundamentals of model-based radio localization and sensing, and is organized as follows. First, the problem definitions, signal and channel models are detailed. Second, relevant performance metrics and bounds are described. Finally,  an overview of the typical methods for localization and sensing is detailed. {In Part II, a complementary literature review is provided, focusing on 6G and its challenges.}

\vspace{-3mm}
\section{Models and Problem Definitions} \label{sec:system model}
In this section, we provide the basic formulations for the localization and sensing problem, within a mobile radio communication context. 
We start with a generic channel model, focusing  on 
{a frequency domain representation with $N$ samples spaced $\Delta_f$ apart, spanning a total bandwidth of $W=N\Delta f$. This representation appears naturally with \ac{OFDM} signals, but it is not limited to them. }

\vspace{-3mm}
\subsection{Generic Observation Model}
\subsubsection{Channel Model}
The channel between  a \ac{Tx} {with $N_{\text{Tx}}$ antennas} and a \ac{Rx} {with $N_{\text{Rx}}$ antennas} over frequency $n \in \{0,\ldots,N-1\}$ and {symbol} 
$k \in \{0,\ldots,K-1\}$ can be approximated by~\cite{heath2016overview}
\begin{align}
    \bm{H}_{n,k} = \sum_{l=1}^{L} \alpha_l \bm{a}_{\text{rx}}(\bm{\theta}_{l})\bm{a}_{\text{tx}}^\top(\bm{\phi}_{l})e^{-\jmath 2 \pi n  \Delta_f \tau_{l}}  e^{\jmath 2 \pi k T_s  \nu_l}, \label{eq:ChannelGeneric}
\end{align}
{where $L$ is the number of physical  propagation paths (as, e.g., would be given by a ray-tracer), $\alpha$ is a complex channel gain, $\bm{a}_{\text{rx}}(\bm{\theta}) \in \mathbb{C}^{N_{\text{Rx}}}$ is the \ac{Rx} array response as a function of the \ac{AoA} $\bm{\theta}\in \mathbb{R}^2$ in azimuth and elevation, $\bm{a}_{\text{tx}}(\bm{\phi})\in \mathbb{C}^{N_{\text{Tx}}}$ is the \ac{Tx} array response as a function of the \ac{AoD} $\bm{\phi}\in \mathbb{R}^2$ in azimuth and elevation, $\tau$ is the \ac{ToA},  $\nu$ is the Doppler shift, and $T_s$ is the 
symbol duration. 
The \ac{AoA} is defined in the reference frame of the \ac{Rx}, the \ac{AoD} in the reference frame of the \ac{Tx}, and thus these angles depend on the respective orientations. Below 6 GHz, due to limited delay and angle resolution, combined with a weak connection of the paths to the environment geometry,\footnote{{Due to complex propagation effects, such as material propagation, diffraction, Rayleigh scattering limited shadowing, and multi-bounce scattering.}} explicit geometric information in the channel is hard to harness. In contrast, at mmWave and above,  paths are more closely related to the environment geometry and 
can be more easily resolved~\cite{HighAccuracyLocalizationforAssistedLiving--Witrisal_others}. Hence, we will assume each path in \eqref{eq:ChannelGeneric}  corresponds to a physical object.}

\subsubsection{Signal Model}
The observation at the \ac{Rx} is then of the form~\cite{heath2016overview}
\begin{align}
    \bm{y}_{n,k} = \bm{W}_k^{\mathsf{H}}\bm{H}_{n,k} \bm{f}_{n,k} + \bm{n}_{n,k}, \label{eq:ModelGeneric}
\end{align}
where  {$\bm{W}_k \in \mathbb{C}^{N_{\text{Rx}} \times M_{\text{Rx}}}$ is an orthonormal} analog \ac{Rx} combiner, with $\bm{W}^{\mathsf{H}}_k \bm{W}_k =\bm{I}_{M_{\text{Rx}}}$ using {$M_{\text{Rx}}\le N_{\text{Rx}}$} RF chains, $\bm{f}_{n,k} $ is the $k$-th \ac{Tx} signal across the \ac{Tx} array, with $\mathbb{E}\{\Vert \bm{f}_{n,k} \Vert^2\}=P_{\text{tx}}/W$, and $\bm{n}_{n,k}\sim \mathcal{CN}(\bm{0},N_0 \bm{I}_{M_{\text{Rx}}})$ is noise after the combining. Here, $P_{\text{tx}}$ is the average transmit power and  $N_0$ denotes the noise power spectral density. {The transmit signals $\bm{f}_{n,k}$ are generally known (pilots in localization or bistatic sensing or known data in monostatic sensing), but may be partially unknown for semi-blind  estimation \cite{wang2015semiblind}.}
\vspace{-3mm}
\subsection{The Localization Problem}
\begin{figure}
    \centering
    \includegraphics[width=1\columnwidth]{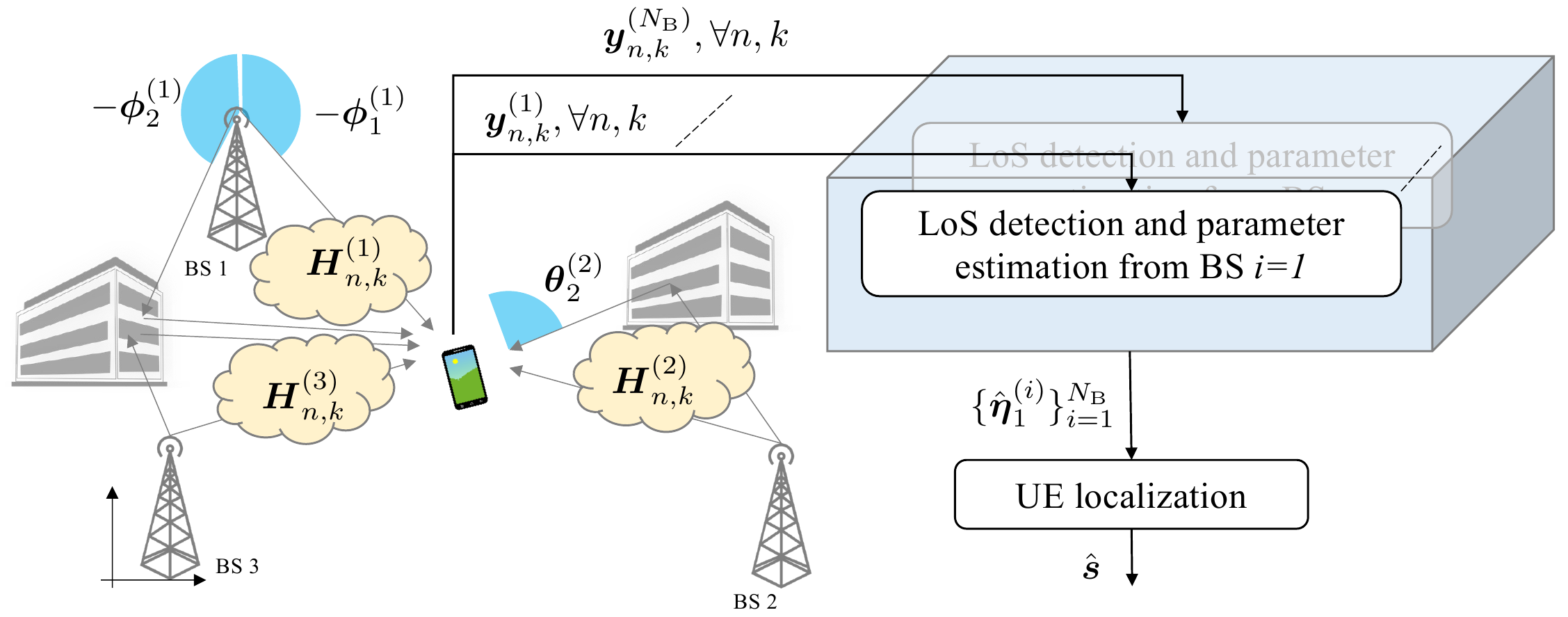}
    \vspace{-8mm}
    \caption{Localization (here shown in \ac{DL}) involves estimating the  \ac{LoS} channel parameters, denoted by $\bm{\eta}^{(i)}_1$, from each BS $i$. }
    \label{fig:positioningdetails}
    \vspace{-5mm}
\end{figure}

In localization, as shown in Fig.~\ref{fig:positioningdetails}, the \ac{UE} has an unknown state $\bm{s}$, which should be inferred from observations of the form \eqref{eq:ModelGeneric}. The state
comprises the position $\bm{x}\in \mathbb{R}^3$, the clock bias ${B}\in \mathbb{R}$, and possibly the orientation $\bm{o}\in \mathbb{R}^3$,  which is often described with over-parameterized representation, such as a rotation matrix $\bm{R}\in \mathbb{R}^{3 \times 3}$ subject to $\bm{R}^\top \bm{R}=\bm{I}$ and $\det(\bm{R})=1$, or a quaternion $\bm{q}\in \mathbb{R}^4$ subject to $\Vert \bm{q}\Vert^2=1$~\cite{barfoot2017state}. 
The infrastructure nodes (\acp{BS} $i\in\{1,\ldots,N_{\text{B}}\}$) have known states, i.e., position $\bm{x}^{(i)}$ and orientation $\bm{o}^{(i)}$, and are time synchronized. Localization can be  user-centric\footnote{{In principle, user-centric monostatic localization is possible without any \ac{BS}, using a priori environment information. See Part II.} } in \ac{DL} or network-centric in \ac{UL}. In \ac{DL}, each \ac{BS} $i$ transmits 
signals over orthogonal subcarriers, leading to observations $\bm{y}^{(i)}_{k,n}$ at the \ac{UE} \ac{Rx} over channels $\bm{H}^{(i)}_{n,k}$, where $i$ has been added to make the BS index explicit. In \ac{UL}, the \ac{UE} transmits a signal, which  reaches the \ac{BS} \ac{Rx} $i$. Note that 
under time division duplexing, 
UL and DL channels are each other's transpose.\footnote{Note also the transpose in \eqref{eq:ChannelGeneric} rather than the Hermitian, typically used in communication.} 

\subsubsection*{Channel Decomposition} \label{sec:ChannelLoc}
While the model \eqref{eq:ModelGeneric} is generic and also widely used in communications, the localization aspect is revealed when we consider the channel $\bm{H}^{(i)}_{n,k}$  and break it up into the \acf{LoS} path and the \ac{NLoS} paths: $ \bm{H}^{(i)}_{n,k}   =\bm{H}^{(i)}_{n,k,\text{LoS}}  + \bm{H}^{(i)}_{n,k,\text{NLoS}}$. 
The \ac{LoS} path  $  \bm{H}^{(i)}_{n,k,\text{LoS}}  = \alpha^{(i)}_{1} \bm{a}_{\text{rx}}(\bm{\theta}^{(i)}_{1})\bm{a}^\top_{\text{tx}}(\bm{\phi}^{(i)}_{1})e^{-\jmath 2 \pi n  \Delta_f \tau^{(i)}_{1}}e^{\jmath 2 \pi k T_s  \nu^{(i)}_{1}}$ (if it is visible and  resolvable) contains geometric information related to the \ac{UE} state $\bm{s}$, via the parameters $\bm{\eta}^{(i)}_1=[(\bm{\theta}^{(i)}_{1})^\top,(\bm{\phi}^{(i)}_{1})^\top,{\tau}^{(i)}_{1},{\nu}^{(i)}_{1}]^\top$. 
The impact of non-resolvable \ac{LoS} is discussed in~\cite{aditya2018survey} and of non-visible \ac{LoS} in \cite{PositionOrientationEstimation--A.Shahmansoori_others_G.Seco-Granados_H.Wymeersch}.
The geometric information brought by each component of $\bm{\eta}^{(i)}_1$ and by $ \alpha^{(i)}_{1}$ is now detailed.
\begin{itemize}[noitemsep,topsep=0pt]
    \item  \emph{\ac{LoS} complex gain $ \alpha^{(i)}_{1}$:} Since the phase of $\alpha^{(i)}_{1}$ varies $2 \pi$ for every movement over one wavelength, it is challenging to account for this (more on this in Part II). The power of $\alpha^{(i)}_{1}$ can be determined by the path loss equation\footnote{{The path loss exponent relates to the \ac{LoS} path only, not any average behavior, as in usual communication channel models.}}
\begin{align}
    \big\vert \alpha^{(i)}_{1}\big\vert^2 = \frac{\lambda^2}{(4 \pi)^2} \frac{ G_{\text{rx}}(\bm{\theta}^{(i)}_{1})G_{\text{tx}}(\bm{\phi}^{(i)}_{1})}{ \Vert\bm{x}-\bm{x}^{(i)} \Vert^2}, \label{eq:channelGainLocalization}
\end{align}
where $\lambda$ is the wavelength at the carrier and $G_{\text{tx}}(\cdot)$ and $G_{\text{rx}}(\cdot)$ denote the antenna element response at the \ac{Tx} and \ac{Rx}, respectively. 
Since these element responses are often only partially known and are affected by environmental variations, the dependence of $\vert \alpha^{(i)}_{1}\vert^2$ on the distance $\Vert\bm{x}-\bm{x}^{(i)} \Vert$ is usually not utilized in the development of algorithms, except for fingerprinting~\cite{vo2015survey}.
\item \emph{\ac{LoS} \ac{AoA} $\bm{\theta}^{(i)}_{1}$ and \ac{AoD} $\bm{\phi}^{(i)}_{1}$:} In \ac{DL}, $\bm{\theta}^{(i)}_{1}(\bm{x},\bm{o})$ is a function of the \ac{UE} position and orientation, while $\bm{\phi}^{(i)}_{1}(\bm{x})$ only depends on the \ac{UE} position. This means that DL-AoA can only be used when the UE orientation is either known or also estimated as part of the state. In \ac{UL}, these dependencies are reversed. The specific expressions depend on how the coordinate systems and angles are defined. For examples, see~\cite{ErrorBoundsfor3DLocalization--Z.Abu-Shaban_others_G.Seco-Granados_H.Wymeersch} or~\cite[Appendix A]{ge20205g}.
  \item  \emph{\ac{LoS} delay $ \tau^{(i)}_{1}$:} It is given by $\tau^{(i)}_{1}(\bm{x}) = \Vert\bm{x}-\bm{x}^{(i)} \Vert/c + B$,  
where $c$ is the speed of light. The clock bias $B$ is due to the lack of synchronization between the BSs and UE and may drift over time. While setting $B$ to a known value is occasionally an assumption in academic papers on localization, it is  overly optimistic and leads to misleading designs and results. 
Dealing with the clock bias can be avoided by using \ac{RTT} measurements{, but should otherwise be estimated as part of the \ac{UE} state.}
  \item  \emph{\ac{LoS} Doppler $ \nu^{(i)}_{1}$:} Due to short transmission times in localization within a coherence interval, $|K T_s  \nu^{(i)}_{1}|\ll 1$, so that Doppler is generally not considered. 
  Nevertheless, Doppler can improve positioning~\cite{han2015performance}, provided it can be disambiguated from the \ac{CFO}. 
\end{itemize}
The \ac{NLoS} channel $\bm{H}^{(i)}_{n,k,\text{NLoS}}$ contains all the multipath, both specular and diffuse, which is traditionally considered as a disturbance. {More on the geometric nature of the \ac{NLoS} channel, the ability to modify the \ac{NLoS} channel via \ac{RIS} \cite{basar2019wireless}, and how to harness multipath is deferred to Part II.} 

\vspace{-3mm}
\subsection{The Sensing Problem}
We consider several point objects (for extended objects, see, e.g.,~\cite{granstrom2016extended}) with state $\bm{s}_l$, including position $\bm{x}_l$ and velocity $\bm{v}_l$ for object $l$. 
An important difference in sensing compared to localization is that the number of objects is a priori unknown in sensing. 
Moreover, objects may appear and disappear from the sensor's field of view and/or may be occluded, leading to missed detections. In addition, clutter may lead to non-existing objects being detected, leading to false alarms. Hence, sensing combines both  detection and  estimation, while localization is essentially an estimation problem. 
The signal and channel model are again of the form \eqref{eq:ChannelGeneric}--\eqref{eq:ModelGeneric}, but the interpretation of the channel parameters is different, depending on whether the transmitter and receiver are co-located (monostatic sensing, as in automotive radar) or not (bistatic or multistatic sensing), as shown in Fig.~\ref{fig:locvssense}. Localization can be seen as a special case of bistatic sensing, where only the \ac{LoS} path is of interest.

\subsubsection*{Channel decomposition}
In both monostatic and bistatic sensing, the channel $\bm{H}_{n,k}$ is broken up as $ \bm{H}_{n,k}   =\bm{H}^{\text{object}}_{n,k}  + \bm{H}^{\text{clutter}}_{n,k}$, where $\bm{H}^{\text{object}}_{n,k}$ captures the part of the channel related to the objects, while $\bm{H}^{\text{clutter}}_{n,k}$ describes the part of the channel related to clutter, e.g., ground reflections, and is modeled statistically.
The different components in the channel bring the following geometric information per resolvable path.
\begin{itemize}[noitemsep,topsep=0pt]
    \item \emph{Channel gain $\alpha_l$:} For monostatic sensing, due to the two-way propagation, the gain is often much smaller than in \eqref{eq:channelGainLocalization}, and is given by  \cite[Vol.~I, Ch.~2]{richards2010principles}
\begin{align}
   \vert \alpha_l\vert^2 = \frac{\lambda^2 \sigma_{\text{RCS},l} G_{\text{rx}}(\bm{\theta}_{l})G_{\text{tx}}(\bm{\phi}_{l})}{(4 \pi)^3\Vert\bm{x}_l\Vert^4},
\end{align}
where $\Vert\bm{x}_l\Vert$ denotes the distance from the sensor (which is seen as the center of the coordinate system) to the object and 
$\sigma_{\text{RCS},l}$ is the \ac{RCS} of the $l$-th object, which depends on the object type. The \ac{RCS} is expressed in $\text{m}^2$ and can range from $1~\text{m}^2$ for a person to $100~\text{m}^2$ for a car. The very small values of $\vert \alpha_l\vert^2$ are compensated by longer integration times, enabled by including the Doppler shift in the observation model. For bistatic sensing, the model is more involved, see, e.g.,~\cite[Section 2.3]{ge20205g}.
\item \emph{\ac{AoA} $\bm{\theta}_{l}$ and \ac{AoD} $\bm{\phi}_{l}$:} In monostatic sensing, the \ac{AoA} and \ac{AoD}  are identical and depend on the object position $\bm{x}_l$. In bistatic sensing, the \ac{AoA} and \ac{AoD} provide independent information about the object~\cite[Appendix A]{ge20205g}.
\item \emph{\ac{ToA} ${\tau}_{l}$:} For monostatic sensing $\tau_l = 2 \Vert\bm{x}_l \Vert/c$, measured with respect to the sensor, 
while for bistatic sensing $\tau_l = (\Vert\bm{x}_l- \bm{x}_{\text{tx}} \Vert+\Vert\bm{x}_l- \bm{x}_{\text{rx}} \Vert)/c+B$, where $B=0$ when \ac{Tx} and \ac{Rx} are time-synchronized.
\item \emph{Doppler ${\nu}_{l}$:} In monostatic sensing, the Doppler is measured in the reference frame of the sensor and given by $\nu_l=2(\bm{v}_l^\top\bm{u}_l)/\lambda$, where $\bm{u}_l$ is a unit vector pointing from the object to the sensor and $\bm{v}_l$ is the relative velocity. For bistatic sensing, the Doppler 
 depends on both the relative velocity and the unit vectors to the target from \ac{Tx} and \ac{Rx}, as well the \ac{CFO}.
\end{itemize}
\vspace{-3mm}

\section{Performance metrics}

While there are many metrics that are of importance, such as latency (the time between the positioning request and the position being available), availability (the fraction of space or time that the localization and sensing service is available with sufficient accuracy), and  scalability (density of \acp{UE} that can be simultaneously supported), our focus will be on  accuracy and resolution. 
\vspace{-3mm}
\subsection{Accuracy} \label{sec:accuracy}
The main performance metric in localization and sensing is accuracy. Let $\bm{e}$ denote the random  estimation error, e.g., for localization $\bm{e}=\bm{x}-\hat{\bm{x}}$, and  $\Vert \bm{e}\Vert^2$ be the $\ell-2$ error norm. Based on percentile or mean values of the error norm, the accuracy is determined, e.g., the \ac{RMSE} or the 90\% percentile. 
For unbiased estimators with $\mathbb{E}\{\bm{e}\}=\bm{0}$  the \ac{RMSE} can,  under certain conditions, be lower bounded by the \ac{CRB}~\cite{VanTrees}. The \ac{CRB} is a powerful tool not only for benchmarking algorithms and predicting performance, but also for deployment and waveform optimization~\cite{HighAccuracyLocalizationforAssistedLiving--Witrisal_others,ErrorBoundsfor3DLocalization--Z.Abu-Shaban_others_G.Seco-Granados_H.Wymeersch} {and for including prior knowledge \cite{buehrer2018collaborative}}. Combining all the observations \eqref{eq:ModelGeneric} yields a long vector $\bm{y}=[\bm{y}^\top_{1,1},\ldots, \bm{y}^\top_{N,K}]^\top$, 
which depends on parameters of interest $\bm{\eta}$ (e.g., the \ac{UE} location) of length $d_{\bm{\eta}}$, and nuisance parameters, say, $\bm{\xi}$ (e.g., channel gains and clock bias). Then, the \ac{FIM} $\bm{J}(\bm{\kappa})$ of $\bm{\kappa}=[\bm{\eta}^\top,\bm{\xi}^\top]^\top$ has as elements\footnote{In most practical cases, the expression is simplified by using the Slepian-Bangs formula 
$\bm{J}(\bm{\kappa})={2}/{N_0}\sum_{n,k} \Re \big\{ \big(\frac{\partial \bm{\mu}_{n,k}}{\partial \bm{\kappa}} \big)^{\mathsf{H}}\frac{\partial \bm{\mu}_{n,k}}{\partial \bm{\kappa}}\big\}$, where  $\bm{\mu}_{n,k}$ is the noise-free observation in $\bm{y}_{n,k}$. } 
\begin{align}
    [\bm{J}(\bm{\kappa})]_{\iota,\iota'}=-\mathbb{E} \Big\{ \frac{\partial \log p (\bm{y}|\bm{\kappa})}{\partial [\bm{\kappa}]_{\iota}} \frac{\partial \log p (\bm{y}|\bm{\kappa})}{\partial [\bm{\kappa}]_{\iota'}}\Big\},\label{eq:FIM}
\end{align}
{where $\log p (\bm{y}|\bm{\kappa})$ is the log-likelihood and $\mathbb{E}\{\cdot\}$ indicates the expectation over the noise.} The following inequality holds:
\begin{align}
    \sqrt{\mathbb{E}\left\{ \Vert \bm{\eta}-\hat{\bm{\eta}}\Vert^2 \right\}} \ge \sqrt{\text{trace}[\bm{J}^{-1}(\bm{\kappa})]_{1:d_{\bm{\eta}},1:d_{\bm{\eta}}}}, \label{eq:CRB}
\end{align}
where the square root is used for easier interpretation of the numerical values. 
When $\bm{\eta}$ is the position, orientation, or velocity, the right-hand side of \eqref{eq:CRB} is known as the \ac{PEB} (expressed in meters), \ac{OEB}, or \ac{VEB}, respectively. We denote $\bm{\Sigma}(\bm{\eta})=[\bm{J}^{-1}(\bm{\kappa})]_{1:d_{\bm{\eta}},1:d_{\bm{\eta}}}$ as the covariance of the estimation of $\bm{\eta}$. 
In certain cases, the FIM lends itself to analytical manipulation, providing deep insights into the nature of the performance due to various factors, such as sensor deployment or bandwidth~\cite{HighAccuracyLocalizationforAssistedLiving--Witrisal_others}. When the FIM is not invertible, the 
problem is non-identifiable{, i.e., there are infinitely many solutions based on the measurements.} 
\vspace{-3mm}
\subsection{Resolution} \label{sec:resolution}
Resolution is the ability to separate  correlated signals. For example, if two objects, say, $l$ and $l'$, in \eqref{eq:ChannelGeneric} have similar \ac{AoA}, \ac{AoD}, \ac{ToA}, and Doppler, they would appear to the receiver as a single object 
with complex gain $\alpha_l + \alpha_{l'}$. Resolution in one domain is sufficient for  objects to be separable. Resolution is also applicable to the localization problem \eqref{eq:ModelGeneric}, to separate the \ac{LoS} path from the \ac{NLoS} channel. We have several domains of resolution~\cite{bilik2019rise}.
\begin{itemize}
    \item \emph{Delay resolution:} Two objects can be resolved if their delay difference $|\tau_l-\tau_{l'}|$ is greater than $1/W$. Hence, a larger bandwidth leads to better delay resolution. For example, 400 MHz of bandwidth leads to a distance resolution of 75 cm.
    \item \emph{Doppler resolution:} Two objects can be resolved if their Doppler difference  $|\nu_l-\nu_{l'}|$ is greater than $1/(KT_s)$. Hence, a larger coherent integration time leads to better Doppler resolution.  For example, to obtain a radial velocity resolution of $1~\text{m}/\text{s}$ with a signal at 30 GHz carrier, the integration time should be at least 10 ms. 
    \item \emph{Angular resolution:} Two objects can be resolved if the difference in azimuth (a similar argument holds for elevation) angle $|\theta^{\text{az}}_l-\theta^{\text{az}}_{l'}|$ is greater than (approximately, since the exact expression depends on the angle itself) $2/N^{\text{az}}$, where $N^{\text{az}}$ is the number of $\lambda/2$-spaced elements along the axis with respect to which the azimuth angle is computed. Hence, larger arrays lead to better angle resolution. For example, to obtain 5 degree angular resolution at boresight, around 23 antennas are required. 
\end{itemize}

\begin{figure}
\centering
%
%
\definecolor{mycolor1}{rgb}{0.92900,0.69400,0.12500}%
\definecolor{mycolor2}{rgb}{0.49400,0.18400,0.55600}%
\definecolor{mycolor3}{rgb}{0.46600,0.67400,0.18800}%
\definecolor{mycolor4}{rgb}{0.30100,0.74500,0.93300}%

\begin{tikzpicture}[scale=1\columnwidth/10cm,font=\footnotesize]
\begin{axis}[%
width=8cm,
height=4cm,
scale only axis,
xmode=log,
xmin=1,
xmax=122.88,
xminorticks=true,
xlabel style={font=\color{white!15!black}},
xlabel={Bandwidth [MHz]},
ymode=log,
ymin=0.001,
ymax=1000,
yminorticks=true,
ylabel style={font=\color{white!15!black}},
ylabel={RMSE of first object [m]},
axis background/.style={fill=white},
 legend columns=3, 
legend style={
  fill opacity=1,
  draw opacity=1,
  text opacity=1,
  at={(-0.01, 0.05)},
  align=left,
  legend cell align=left,
  anchor=south west
}
]
\addplot [color=red, dashed, line width=2.0pt,mark=*]
  table[row sep=crcr]{%
0.96	10.1509694172485\\
1.92	5.14970161832234\\
3.84	2.5408092751538\\
7.68	1.30674528931586\\
15.36	0.652093111554356\\
30.72	0.340189886532389\\
61.44	0.164912997200748\\
122.88	0.0802661635651633\\
};
\addlegendentry{RMSE (1 path)}

\addplot [color=blue, dashed, line width=2.0pt]
  table[row sep=crcr]{%
0.96	58.763017051396\\
1.92	46.1427504108384\\
3.84	38.7833344095675\\
7.68	29.706074064325\\
15.36	7.57583824994693\\
30.72	0.8822787811436\\
61.44	0.22034914153635\\
122.88	0.0900872668620709\\
};
\addlegendentry{RMSE (5 paths)}

\addplot [color=mycolor1, line width=2.0pt]
  table[row sep=crcr]{%
0.96	312.5\\
1.92	156.25\\
3.84	78.125\\
7.68	39.0625\\
15.36	19.53125\\
30.72	9.765625\\
61.44	4.8828125\\
122.88	2.44140625\\
};
\addlegendentry{resolution}

\addplot [color=black, line width=1.0pt]
  table[row sep=crcr]{%
0.96	20\\
1.92	20\\
3.84	20\\
7.68	20\\
15.36	20\\
30.72	20\\
61.44	20\\
122.88	20\\
};
\addlegendentry{inter-path distance}

\addplot [color=red, line width=2.0pt]
  table[row sep=crcr]{%
0.96	10.9827344826809\\
1.92	5.45896951173931\\
3.84	2.72547964035069\\
7.68	1.36224055662308\\
15.36	0.68105791322618\\
30.72	0.340521162315681\\
61.44	0.170259606912477\\
122.88	0.0851296816768743\\
};
\addlegendentry{CRB (1 path)}

\addplot [color=blue, line width=2.0pt]
  table[row sep=crcr]{%
0.96	14611590.9303994\\
1.92	3759809.28196941\\
3.84	24833.2707559492\\
7.68	229.619157136688\\
15.36	2.89520309818785\\
30.72	0.364853998550178\\
61.44	0.172989796619058\\
122.88	0.0854820728286714\\
};
\addlegendentry{CRB (5 paths)}

\end{axis}

\end{tikzpicture}%
\vspace*{-1cm}
\caption{An evaluation of the CRB (discussed in Section \ref{sec:casestudy}) and  estimation RMSE  (discussed in Section \ref{sec:estimation}) of the closest among 5 objects as a function of the signal bandwidth. Here, `1 path' refers to the case with only one object. }
\label{fig:sim_vs_crb}\vspace{-5mm}
\end{figure}
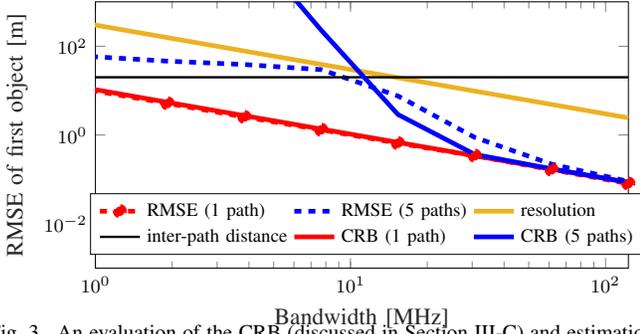
\vspace{-3mm}
\subsection{Case Study} \label{sec:casestudy}
To show that resolution is needed to achieve high accuracy, 
it is instructive to consider a simple example. In Fig.~\ref{fig:sim_vs_crb}, a single-antenna bistatic sensing scenario with 5 objects is evaluated, with inter-object spacing of 20 meters,  each with same channel magnitude, set to achieve a  10 dB SNR per object. The figure shows  the RMSE of the first object as a function of bandwidth. First, consider the inter-path spacing (black) and resolution (orange). When the resolution curve is below the inter-path spacing (this happens at about 20 MHz), we expect the objects to become resolved. This is confirmed by the \ac{CRB}: the CRB (blue curve) is high for low bandwidths and around 20 MHz starts to reach the CRB of the single-object case (red curve). 
Hence, performance far better than the waveform resolution is possible, e.g., by increasing the integrated SNR, provided the signal paths can be resolved. Without sufficient resolution, accuracy is limited. 

\vspace{-4mm}
\section{Localization and Sensing Methods}

From the observations $\bm{y}_{n,k}$, it would be tempting to solve the localization or sensing problem by direct optimization. Letting $\bm{\kappa}$ contain the \ac{UE} state (for localization) or object states (for sensing) as well as any nuisance parameters (channel gains), then 
\begin{align}
    \hat{\bm{\kappa}}= \arg \max_{\bm{\kappa}} \log p(\bm{y}|\bm{\kappa}),
\end{align}
where the relevant quantities were already defined in Section \ref{sec:accuracy}. This is known as \emph{direct positioning} and has as benefit that all possible information is used, though at a high computational cost~\cite{DirectLocforMassiveMIMO--N.Garcia_H.Wymeersch_R.Larsson_others}. Most practical methods apply a two-stage approach, already hinted at in Fig.~\ref{fig:positioningdetails}, whereby first the geometric channel parameters (angles, delays, Dopplers) are estimated, and then the UE / object state is recovered. 

\subsection{Channel Parameter Estimation} \label{sec:estimation}
As the channel estimation problem is also present in wireless communication, there exists a variety of estimators, including FFT/Periodograms~\cite{braun2014ofdm}, ESPRIT~\cite{roemer2014analytical}, generalized approximate message passing~\cite{bellili2019generalized}, orthogonal matching pursuit~\cite{PositionOrientationEstimation--A.Shahmansoori_others_G.Seco-Granados_H.Wymeersch}, and RIMAX/SAGE~\cite{thoma2004rimax}, which exploit either underlying sparsity or principles of harmonic retrieval (or both). 
A common approach is to first obtain an unstructured estimate 
$\hat{\bm{H}}_{n,k}$ of the channel \eqref{eq:ChannelGeneric} from a \ac{LS} estimator. 
Introducing $\bm{a}_{\text{d}}(\tau)=[1, \ldots, e^{-\jmath 2 \pi (N-1)  \Delta_f \tau_{l}}]^\top$ and $\bm{a}_{\text{D}}(\nu)=[1, \ldots, e^{\jmath 2 \pi (K-1) T_s  \nu_l}]^\top$, after vectorizing  and stacking  of $\hat{\bm{H}}_{n,k}$, we can write\footnote{{The vectors $\bm{a}_{\text{tx}}(\bm{\phi})$ and $\bm{a}_{\text{D}}(\nu)$ are coupled, which can be resolved by imposing additional assumptions~\cite{cheng2012joint}.}}
\begin{align}
    \hat{\bm{h}}=\sum_{l=1}^L \alpha_l \bm{a}_{\text{d}}(\tau_l) \otimes \bm{a}_{\text{D}}(\nu_l) \otimes  \bm{a}_{\text{rx}}(\bm{\theta}_{l}) \otimes \bm{a}_{\text{tx}}(\bm{\phi}_{l})+\bm{n}, \label{eq:vectorizedObs}
\end{align}
which is in an appropriate form for compressive sensing methods~\cite{lee2016channel}. This was applied in Fig.~\ref{fig:sim_vs_crb}, where orthogonal matching pursuit was used to detect the number of paths and estimate their delays. When the resolution is above the inter-path distance, the number of detected paths is too few, leading to biased estimates, which explains the RMSE (dashed) below the CRB (in blue). When the resolution is below the inter-path distance, the RMSE follows the correspond CRB quite well, and then attains the RMSE for the single-path case. 

Alternatively, we can express the \ac{LS} estimates in a tensor form \vspace{-3mm}
\begin{align}
    \hat{h}_{i_1,i_2,\ldots, i_D} = \sum_{l=1}^{L} \alpha_l \prod_{d=1}^{D} e^{\jmath i_d \omega_{d,l}} + n_{i_1,i_2,\ldots, i_D}, \label{eq:HRproblem}
\end{align}
where $\omega_{d,l}$ is a so-called spatial frequency. For instance, if index $d$ refers to the subcarrier dimension, then $\omega_{d,l}=- 2 \pi  \Delta_f \tau_{l}$. Now, \eqref{eq:HRproblem} is a classic harmonic retrieval problem in $D$ dimensions ($D=4$ in \eqref{eq:vectorizedObs} but can be expanded to $D=6$ if the \ac{Tx} and \ac{Rx} array admit a Kronecker structure). 
Once the number of objects $\hat{L}$, their gain $\hat{\alpha}_l$ and their geometric parameters   $\hat{\bm{\eta}}_l=[\hat{\bm{\theta}}_{l}^\top,\hat{\bm{\phi}}_{l}^\top,\hat{\tau}_{l},\hat{\nu}_{l}]^\top$ have been estimated, they can be further refined by optimization of the log-likelihood function $\log p(\bm{y}|\bm{\kappa})$ around this initial estimate. If the initial estimate is good enough, this will lead to an efficient estimate, close to the CRB with inverse FIM, say $\bm{\Sigma}(\hat{\bm{\eta}}_l)$, which can be used as an uncertainty estimate.  

\vspace{-3mm}

\subsection{Position Estimation}

Estimating the state of a UE or an object now relies on the relationship between the estimated channel parameters, expressed as $\hat{\bm{\eta}}$ and the corresponding uncertainty $\bm{\Sigma}(\hat{\bm{\eta}})$, to the state of interest. 
We focus on the localization problem for concreteness.\footnote{{For sensing, a similar process is performed for each detected target.}} Starting from $\hat{\bm{\eta}}^{(i)}_1=[(\hat{\bm{\theta}}^{(i)}_{1})^\top,(\hat{\bm{\phi}}^{(i)}_{1})^\top,\hat{\tau}^{(i)}_{1},\hat{\nu}^{(i)}_{1}]^\top$ and associated uncertainty $\bm{\Sigma}(\hat{\bm{\eta}}^{(i)}_1)$ of the \ac{LoS} path from each BS $i$ (see Fig.~\ref{fig:positioningdetails}), the \ac{UE} state is related to these channel parameters measurements through $\hat{\bm{\eta}}^{(i)}_1 = \bm{h}^{(i)}(\bm{s}) + \bm{n}^{(i)}$,  
where $\bm{n}^{(i)} \sim \mathcal{N}(\bm{0},\bm{\Sigma}(\hat{\bm{\eta}}^{(i)}_1))$ and $\bm{h}^{(i)}(\bm{s})$ is the known mapping from \ac{UE} state to the geometric channel parameters, as described in Section \ref{sec:ChannelLoc}. This enables us to express the problem\footnote{{The Gaussian model is justified through the extended invariance principle~\cite[Thm.~1]{swindlehurst1998maximum}.}}
\begin{align}
     \hat{\bm{s}}=\arg \min_{\bm{s}} 
     \sum_{i=1}^{N_{\text{B}}}  \big( \hat{\bm{\eta}}^{(i)}_1 - \bm{h}^{(i)}(\bm{s})\big)^\top{\bm{\Sigma}^{-1}(\hat{\bm{\eta}}^{(i)}_1)}\big( \hat{\bm{\eta}}^{(i)}_1 - \bm{h}^{(i)}(\bm{s})\big) \notag 
\end{align}
which is non-convex and can be solved by first obtaining  a coarse estimate  (e.g., using geometric reasoning, linearization, or relaxation). It is then refined by local optimization of the likelihood function. 
The final estimate $\hat{\bm{s}}$ is then used to compute a covariance $\bm{\Sigma}(\hat{\bm{s}})$ from 
\begin{align}
    \bm{\Sigma}^{-1}(\hat{\bm{s}}) =\left. \sum_{i=1}^{N_{\text{B}}}
    \Big(\frac{\partial {\bm{\eta}}^{(i)}_1}{\partial \bm{s}}\Big)^\top\,  {\bm{\Sigma}^{-1}(\hat{\bm{\eta}}^{(i)}_1)}   \frac{\partial {\bm{\eta}}^{(i)}_1}{\partial \bm{s}}\right|_{\bm{s}=\hat{\bm{s}}}.
\end{align}
The couple $(\hat{\bm{s}},\bm{\Sigma}(\hat{\bm{s}}))$ can then be further processed, e.g., in a tracking filter or sensor fusion engine. 

\vspace{-3mm}
\section{Conclusions}
In this letter, we have provided an overview of the radio localization and sensing problems, described the basic models, performance metrics and methods. An important focus was on modeling of channels and signals, which is needed to develop practical methods with high accuracy and reasonable uncertainty information. We also emphasized the need for high resolution as a prerequisite for high accuracy. This overview provides the  background for the more advanced principles in Part II. 
\vspace{-3mm}
\bibliographystyle{IEEEtran}
\bibliography{references.bib} 
\end{document}